\documentstyle[sprocl]{article}

\def\be{\begin{equation}}
\def\ee{\end{equation}}
\def\bea{\begin{eqnarray}}
\def\eea{\end{eqnarray}}

\begin{document}

\title{NEW REGIME FOR DENSE STRING NETWORKS\footnote{
Talk presented at COSMO99, Trieste,27 Sept. -2 October 1999, based on work done in
collaboration with M. Nagasawa and V.Soni} }

\author{Charanjit S. Aulakh}

\address{Department of Physics, Panjab University,\\
Chandigarh, INDIA\\E-mail: aulakh\%phys@puniv.chd.nic.in}

 
\maketitle\abstracts{ 
We uncover a new transient regime that reconciles the apparent
inconsistency of the Martins Shellard one scale damped string
evolution model\cite{ms} with the initial conditions predicted by the
Kibble mechanism for string formation in a second order phase
transition. This regime carries (in a short cosmic time $\sim .1
t_c$) the dense string network created by the
Kibble {\it{mechanism}} to the (dilute) Kibble {\it{regime}} 
 in which friction dominated strings remain
till times $t_* \sim (M_P/T_c)^2 t_c$.
This is possible beacause the cosmic time at the phase
transition ($t_c$) is much larger than the damping time scale
 $l_f\sim T_c^2/T^3$.
Our result has drastic implications for various non-GUT scale string mediated
mechanisms. }

\section{Introduction}

Topological strings are an inevitable  consequence of symmetry
breaking phase transitions when the first homotopy group of
the vacuum manifold is non trivial. They can trap gauge flux 
and could have  have important consequences for the structure
of the present day universe\cite{topdef}.

Most research has focussed on strings formed at temperatures
above GUT scales (which are relevant for galaxy
formation. Recently mechanisms for baryogenisis, neutralino
production etc\cite{mech}  based on dense
string networks formed in second order phase transitions at
scales far below the GUT scale have also been proposed. Such
networks would be friction dominated till very low teperatures
$t_* \sim (M_P/T_c)^2 t_c$ . An estimate
 of how the network correlation length $L$ evolves between  is a
critical input for such models. 

Due to Everett  and Aharanov-Bohm scattering\cite{topdef} a string with
velocity $\vec{v}$ relative to the ambient plasma feels
a velocity dependent frictional force characterized
by a lengthscale $l_f=\eta^2/\beta T^3$ ($\beta$
is a numerical parameter $\sim 10^{-2}- 1$). String curvature causes
continual motion so energy conservation implies loss of string
length due to frictional damping . For non-GUT
strings this is the dominant cause of dissipation of string
length over much of the network's history.  Thus  any 
damped network evolution model must account for the 
average string velocity $v$ . Such a simple one scale model which
tracks both $L$ and $v$ was given by  Martins and Shellard(MS)\cite{ms}.
Their equations read :

\begin{eqnarray}
{dL\over dt}=& HL(1 + v^2) + \displaystyle{\frac{L v^2}{2l_f}}
+\displaystyle{\frac{c v}{2}}\ , \nonumber \\ 
{{dv \over dt}}=& (1-v^2)\left[ \displaystyle{\frac{k}{L}}-v\left( 2 H +
\displaystyle{\frac{1}{l_f}}\right)\right]\ ,
\label{lvt}
\end{eqnarray}
where $H$ is the Hubble parameter, $k$ characterizes the
``wiggliness'' and c the self intersection or ``chopping'' rate.
$k$ is $\sim 1$ in the
frictional era but smaller at the late times of the
linear scaling regime when the strings are undamped and therefore  kinky .
For dense networks reconnection of loops back  may be important
leading to a lower effective value of $c$ but geometrical
estimates give $c\sim 1$.

It is important that $l_f(t_c) \sim \eta^{-1} <<t_c\sim
M_P/\eta^2$  when ($\eta << M_P$). Thus a short interval of cosmic time
is very long in units of the frictional length scale.{\it{ This
simple observation is at the root of the new regime proposed in
this paper}}.

In the MS model the  network curvature radius $R\equiv L$.
So the curvature force is never zero which forces the string to
move and dissipate its rest energy . One 
may visualize the growth of $L$ and $R$  as being due
to the decaying away of curved and looped regions with high curvature
leaving behind a network with less string (larger $L$) and lower
curvature (larger $R$).

Clearly the initial correlation length $L_i$ must obey 
$\eta^{-1} < L_i <t_i\sim \sigma/\eta$ where $\sigma \sim M_P/\eta$.
These limits correspond to those expected in a second
and first order transition respectively, while  on energetic grounds
one expects $v_i\sim 1$. 

The Kibble mechanism\cite{topdef}  predicts
$\epsilon_K$ string lengths of size $1/T_c$ are formed in each
correlation volume $\sim T_c^{-3}$ (where $\epsilon_K\sim
10^{-1} - 10^{-3}$) so $L_i\sim \epsilon_K^{-1/2}/T_c$.

MS consider only the range 
${\sqrt\sigma}/\eta < L_i <\sigma/\eta$ corresponding to
initially non-dense networks. They found that the phase
transition is followed by 

i) {\it{Stretching Regime}}: it
occurs only when $L_i > {{\sqrt\sigma}}/\eta$ . As
$L_i\rightarrow {{\sqrt\sigma}}/\eta$ the duration of the 
stretching regime goes to zero. $ L,v$ scale as :

\begin{equation}
L_S= L_i ({t\over t_c})^{1/2} \qquad\qquad   v_S={t\over{\sigma L_i}} 
\label{LS}
\end{equation}

ii) The {\it{Kibble Regime}}\cite{kibreg} follows the
{\it{stretching}} regime . It is a terminal velocity regime
arising from the balance of the frictional and curvature forces.
\begin{equation}
{L_K} = \left(\displaystyle{\frac{2 k (k +c) }{ 3 \sigma}}
\right) ^{1/2} \left(\displaystyle{{t}}\right)^{5/4}{t_c}^{-1/4}\ ; \qquad
v_K= \left(\displaystyle{\frac{3 k }{2 (k+c) \sigma}}\right) ^{1/2}
\left(\displaystyle{\frac{t}{t_c}}\right)^{1/4}\ 
\label{lvkibb}
\end{equation}

The Kibble regime begins at $t_K\sim
L_i^{4/3}M_P^{1/3}$, {\it{i.e earlier lor lower $L_i$}}.
 Since $t_K \geq t_c$ for consistency it appears as if
$L_i\geq\sqrt\sigma /\eta$ necessarily ! Now $L_K(t_c)\sim
\sqrt\sigma /\eta$ while a second order phase transition
predicts $L_i\sim 1/\eta$. There is a wide
mismatch between the two values so it appears the MS
model leads one to the contradictory conclusion that although the Kibble
regime begins immediately after the phase transition yet its
density must necessarily be much lower than that predicted by the
Kibble mechanism . A similar mismatch holds for $v$ .

The resolution lies in the observation that $l_f(t_c)<< t_c$. 
The correct time scale to use when $t\sim t_c$ is $l_f \sim 1/\eta$
not $t_c$. Consider then the  evolution of $L$ and $v$ in the
period immediately following the phase transition i.e when 
${\hat t} = t-t_c' << t_c'$ , where $t_c'\sim  10t_c$ (say).
The precise value of  $t_c'\sim t_c$ (at which the 
evolution of the one scale model is taken to begin) is
unimportant as can be checked
by varying $t_c'$ by a factor of $10$.
We define a dimensionless time $x=\eta \hat t$
and network scale $l=\eta L$. 
Note that this amounts to
measuring time in units of the damping time scale and and avoids
the introduction of large numbers in the evolution
equations which result if one uses $t_c$ as the unit of time.
 Then the
condition  ${\hat t} = t-t_c' << t_c$ translates to $x<< \sigma$.
Even for GUT scale strings $\sigma$ is $\sim
10^4$ so the variable $x$ can change by many orders of magnitude before
the approximation becomes invalid. Expanding the terms in the
evolution equation around $t_c'$ one finds that as long as
 $l<\sigma^{1/2}$ and $v > \sigma^{-1/2}$
the Hubble term is completely dominated by the velocity
dependent terms. To leading order we have 

\begin{equation}
2 { dl\over dx}=  {{l v^2 \beta}} + c v\ ;\qquad
{dv\over dx}= (1-v^2) \left( \displaystyle{\frac{k}{l}}
-{v \beta} \right) .
\label{lvx}
\end{equation}

We must find the behaviour of $l$ and $v$
for large $x$ beginning from  natural initial conditions
(Kibble Mechanism) where $l_i\sim 1$ .  $l$
always increases while $v$ will increase when $k\geq v\beta l$
and vice versa. Regardless of of $v_i$ the evolution 
locks on  to a trajectory where $v$ is very
closely approximated by $k/{\beta l}$. Then for $\sigma>>x>>1$ 
one gets

\begin{equation}
l= \left( \displaystyle{\frac{(k +c) k}{\beta}}\right)^{1/2}x^{1/2}\ ;\qquad
v= \left(\displaystyle{\frac{k}{\beta (k+c)}}\right)^{1/2} x^{-1/2}\ .
\label{lv}
\end{equation}

This regime persists until the growth of the Hubble term and the
decrease of the velocity dependent term makes them comparable
i.e till  $x\sim 10^{-1} \sigma$ (for typical values of $k,c$)
 by when $t\sim  t_c' + .1 t_c$.
After this the effect of the Hubble terms causes a shift in the
power-law exponents and the Kibble regime regime begins.
Our new regime makes it possible to extend the applicability of
the MS model into the regime of initially dense networks where
it appeared to give a contradiction. When it ends one has $L\sim
{\sqrt\sigma}\eta^{-1}, v\sim \sigma^{-1/2}$, which are just
right to match the Kibble regime beginning at $t\sim t_c'$.

The most remarkable feature of this regime is thus that {\it the
values of $L$ and $v $ very shortly after the phase transition
are essentially independent of their initial values}. This
is in sharp contrast to the ``stretching regime '' $L\sim
L_i (t/t_c)^{1/2} ,
v\sim {t (L_i\sigma)^{-1}}$ found 
for initial conditions $ L_i > \sigma^{1/2}/\eta$.

The Lyapunov exponents obtained by
linearizing the eqns. (\ref{lvx}) around the solutions eqns. (\ref{lv}) 
are$\sim -c/2(k+c)x, -\beta$ i.e  negative so that
our regime is an attractor as $x$ increases. 
We have also verified the above asymptotic  analysis given 
by numerical integration.
For large $x $ the asymptotic solutions
given in eqn.(\ref{lv}) are in excellent agreement with the 
results of numerical integration of eqns.(\ref{lvx}) irrespective
of whether the initial velocities are relativistic or very small.

Similar conclusions hold for string length in loops at the time
of formation or that in loops that form by self-intersection.
However superconducting loops (vortons) may evade the effects
of curvature driven damping since the supercurrent on vortons can
stabilize their radius.

To summarize : The one scale model with friction and velocity
evolution\cite{ms} leads to a consistent picture of network
evolution during the entire friction dominated era.  It predicts
that {\it{the dense, fast network formed in a second order phase
transition crosses over into a sparse slow network in a time
$\sim .1t_c$. }}
Thus mechanisms\cite{mech}
 that rely on the high density of string length
predicted by the Kibble mechanism are unlikely to work unless
string loops are stabilized against curvature driven
contraction, e.g by superconducting currents. Our analysis shows
that friction can be a very important mechanism for string
decay. Further analysis of the microscopic mechanism of
frictional conversion of the coherent string condensate into
plasma energy is required.

\end{document}